# Artificial ciliary bundles with nano fiber tip links

Mohsen Asadnia[1,2+], Ajay Giri Prakash Kottapalli[2+], Jianmin Miao[1*], Michael Triantafyllou[2,3]

Mechanosensory ciliary bundles in fishes are the inspiration for carefully engineered artificial flow sensors. We report the development of a new class of ultrasensitive MEMS flow sensors that mimic the intricate morphology of the ciliary bundles, including the stereocilia, tip links, and the cupula, and thereby achieve threshold detection limits that match the biological example. An artificial ciliary bundle is achieved by fabricating closely-spaced arrays of polymer micro-pillars with gradiating heights. Tip links that form the fundamental sensing elements are realized through electrospinning aligned PVDF piezoelectric nano-fibers that link the distal tips of the polymer cilia. An optimized synthesis of hyaluronic acid-methacrylic anhydride hydrogel that results in properties close to the biological cupula, together with drop-casting method are used to form the artificial cupula that encapsulates the ciliary bundle. In testing, fluid drag force causes the ciliary bundle to slide, stretching the flexible nanofibers, thus eliciting charges. Functioning with principles analogous to the ciliary bundles, the sensors achieve a high sensitivity and threshold detection limit of 0.286 mV/(mm/s) and 8.24 μm/s, respectively, and exhibit a direction-dependent output.

[1]School of Mechanical & Aerospace Engineering, Nanyang Technological University, Singapore 639798.
[2]Center for Environmental Sensing and Modeling (CENSAM) IRG, Singapore-MIT Alliance for Research and Technology (SMART) Centre, Singapore 117543. [3]Department of Mechanical Engineering, Massachusetts Institute of Technology, Cambridge, MA 02139.

[+]Authors with equal contribution

# 1. Introduction

Most animal sensory systems work on the principle of mechanotransduction and offer a diverse set of exceptionally sensitive sensors. Nature's evolutionary path led to sensors of high functionality and high robustness, both in terms of the materials used and energy needed. Through the principle of mechanotransduction, the biological sensors serve various sensing needs in amniotic vertebrates including crickets, amphibians, fishes and humans. The basic units are tiny haircell bundles, which translate mechanical stimuli into electrical signals across the cell membrane [1-3]. For instance, haircells on the cricket can detect wind flows as low as 0.01 mm/s, with a bending angle of hair shaft of about $10^{-4}$ radians at 1mm/s flow velocity [4]. Cochlear haircells found in the inner ear respond with high speed to even nanometer scale deflections [5]. Outer ear haircells act as detectors, but also perform a crucial role in enhancing the frequency selection and sensitivity [6,7]. The lateral-line system in fishes and amphibians utilize arrays of haircell sensors for flow velocity and direction sensing [8]. Haircells in the insect limb joints detect the orientation of their angled leg joints [4]. Research has shown that, despite their simplicity and ubiquity, haircell sensors demonstrate outstanding capabilities, such as ultra-high flow sensitivity and very low detection limits, hence combining high resolution and low threshold detection limits [2,9]. The structural and functional similarity between the auditory haircells found in the inner ear of mammals, and the lateral-line haircells found in fishes and amphibians, makes the stereovilli system a prime candidate to be studied and mimicked in order to devise artificial sensors. Most of these haircells consist of bundles of cilia with gradiated heights that protrude into the external flow. Flow variations cause deformation of the cilia, which in turn cause opening of stretch-activated ion channels (figure 1).

Superficial neuromast sensors in blind cavefish are epithelial organs, approximately 10-500 μm in diameter, composed of a number of stereocilia bundles embedded within a gelatinous cupula that mechanically couples external flows to the stereocilia [8,10]. Each sensing bundle consists of a number of stereovilli, which extend from the apical surface into the cupula. Each ciliary bundle consists of a single long kinocilium and several shorter stereovilli located on one side of it. The stereovilli are organized in a step-wise increasing height fashion from the shorter ones (farthest from the kinocilium) to the longest ones (closest to the kinocilium) [11]. The tips of all the stereovilli are connected to each other with tiny "tip links". The tip links connect the stereocilia along the short-to-tall graded height rows with each link extending upwards connecting to its taller neighbour. Stretching of ion channels results in the flow of action potentials in the afferent nerve fibers that connect the sensory periphery to the brain [10]. The tip links stretch or contract during the bending of the haircell bundle depending on the direction of bending. Due to the step-wise increasing height variation of the stereovilli, flows that cause the stereovilli to bend towards the kinocilium cause a stretching of the tip links. Deflection of the kinocilium away from the stereovilli, and towards the stereovilli, elicit an excitory and inhibitory response respectively in the afferent nerves [12]. Superficial neuromasts, owing to their unique structure and organization of the stereocilia bundle demonstrate an excellent ability to sense the direction of flows. The amount of activated stretch in ion channels depends on morphological polarization of the transduction apparatus. Stereovilli bundles within an individual neuromast located next to each other are positioned out-of-phase with respect to each other [10,13]. Therefore, flow in a particular direction causes a maximum sensitivity for one set of ciliary bundles and a minimum sensitivity for the adjacent ones. The response of a single ciliary bundle varies as a cosine function to the incident flow direction with respect to the height-graident organization of the stereovilli [12]. This orthogonal organization of adjacent haircells seems to enhance the directional sensitivity of the neuromast [14]. Each haircell bundle covers for the worst functioning scenarios of the adjacent ones [11]. Through such division of labor the neuromast achieves an enhanced directional sensitivity.

The sensing abilities achieved by biological haircell sensors are beyond the capabilities of most human-engineered sensors available today. Haircell-like sensor design for artificial microelectromechanical system (MEMS) flow sensors offers several advantages. First, it allows imitating novel structural and functional principles found in nature. Second, it allows easily to introduce improvements to simplify the fabrication process to suit the application. As we show, the combination of sensing principles adapted from nature, together with MEMS design and fabrication methods enables the development of devices with sensing abilities that surpass those of conventionally engineered sensors and match the high sensitivity of their biological counterparts.

Engineering a replication of the intricate morphological organization of the haircell is a complex and challenging task. In most past developments, the haircell has been approximated as a cylindrical pillar on a piezoelectric or piezoresistive membrane or cantilever [15-20] to emulate the flow sensing functionality of the superficial neuromasts [21-23]. This paper presents a biomimetic approach of development of artificial superficial neuromasts that closely imitate the material and structural organization of the biological superficial neuromast. Graded rows of flexible bundled polydimethylsiloxane (PDMS) polymer pillars form a similar structure to that of stereovilli in the fish sensory system. The electrospun piezoelectric polyvinylidene fluoride (PVDF) polymer nanofibers connect consecutive tips of the pillars to mimic the biological tip links, which constitute the core sensing elements in the superficial neuromasts. In addition, biomimetic materials with properties similar to those of the biological cupula are characterized through nano-indentation and rheological studies (provided as a supplementary document). The hydrogel micro-structures developed through drop-casting and swelling mimic the biological cupula that is made up of gelatinous glycoprotein material.

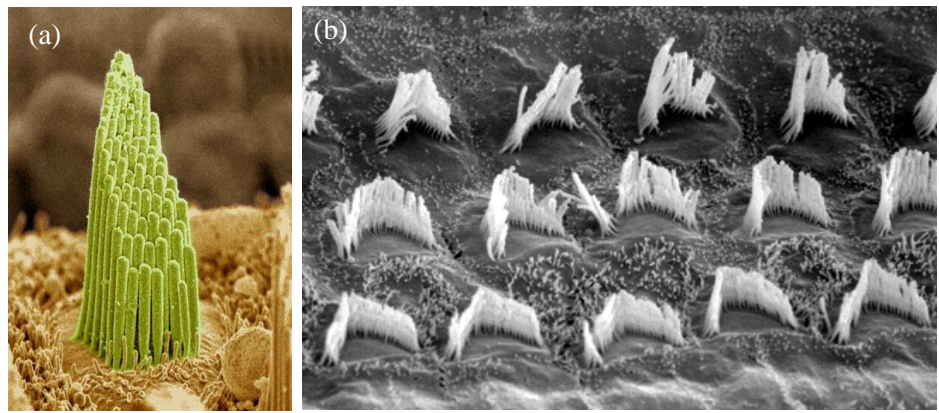

Figure 1. (a) A coloured SEM image of the individual haircell bundle showing the stereovilli organization [9]. (b) An array of ciliary bundles in the inner ear.

**RESULTS AND DISCUSSION**

Figure 2a depicts the general structure of the proposed flow sensor. In this design, each artificial ciliary bundle consists of 55 PDMS stereovilli that are arranged into 10 graded rows. The device features a single tallest pillar (kinocilium) and 54 stereovilli arranged into 10 rows with gradating height in such a way that all the pillars of within each row have the same height. The design follows biomimetic principles to allow the sensor to detect flow velocity and direction, while providing a feasible fabrication process, despite difficulties in achieving such gradients in height. The row of the shortest pillars is farthest from the kinocilium. Each stereocilium of a successive row of higher height falls at the center of the two stereovilli of the preceding shorter row. This arrangement maximizes the number of tip links generated between the pillars and also ensures that all the tip links have similar length, equal to the distance between the rows. This structure and dimensional design of the stereovilli is inspired by studies on the biological stereovilli [13,24]. Each PDMS stereocilium has a diameter of 50 μm and the successive stereovilli are spaced at 25μm distance. The entire distance between the tallest and the shortest stereovilli, located farthest from the kinocilium spans a length of 725μm. The graded height increment of the PDMS pillars ranges from 200μm for the shortest pillar row to 400μm for the tallest pillar (kinocilium) as shown in figure 2b. Elaborate fabrication with details of unit processes is explained in supporting document.

The tip links in the artificial stereovilli sensor are formed through electrospinning piezoelectric PVDF nanofibers. PVDF has been chosen as the sensing element due to its highest piezoelectric coefficient among other polymer materials and its flexibility while it also enables the formation of nanofibers through electrospinning process [17]. PVDF forms in different crystal structures depending on sample preparation conditions. In nature PVDF appears in different phases which are known as α, β, γ and δ [25]. Each phase can be transferred to the other phases under certain external conditions. In general, α-phase is the most commonly available phase in nature, typically obtained when the PVDF is cooled and solidified from melt. While the α-phase is known as a non-polar structure that does not show piezoelectricity, the β-phase is realized to be the only PVDF ferroelectric crystalline structure (polar) with strong piezoelectric effect. In general, high mechanical (approximately 50%) and electrical stretches are required to predominantly convert the PVDF from α-phase to β-phase. In order to determine the molecular and crystalline structure of the electrospun PVDF we performed various characterization experiments such as X-ray Diffraction (XRD), Fourier transform infrared spectroscopy (FTIR) which are presented in supplementary document.

Stretching of the nanofiber causes charges to be generated, due to the piezoelectric nature of the PVDF material, which are acquired as output voltage. These nanofibers form the central part of the MEMS flow sensor. While other components, including the stereovilli and cupula, form the structural design parameters, the nanofibers form its essential sensing component.

The hydrogel cupula is an important part of the proposed flow sensor that not only interacts with the flow and transmits the fluid drag force to the embedded haircells, but also acts as a package protecting the fragile nanofibers [26]. The presence of the hydrogel cupula enhances the sensitivity of the MEMS flow sensor in many ways. The height and diameter of the hydrogel cupula are larger than those of the haircells; hence increasing the surface area projected to the flow which leads in an increased drag force and thereby an enhanced sensitivity. Also, the increased height of the cupula over the haircells causes the structure to extend beyond the boundary layer generated by the body in the flow, further enhancing sensitivity. To develop an artificial cupula, we identified hyaluronic acid-methacrylic anhydride (HA-MA) hydrogel as a material with close properties to the biological cupula [27]. The HA-MA hydrogel material has a density close to the density of water and can be mainly driven through viscous forces [8]. The cupula diminishes the effects of high frequency flows and Brownian motion, thereby increasing the signal to noise ratio of the MEMS flow sensors [28]. The hydrophilicity and permeability of the hydrogel material increase the signal absorption through an enhanced friction factor associated with the material [29,30]. Here, we use an analysis developed by Peleshenko et al. [31] to theoretically estimate the sensitivity enhancement due to the presence of the cupula over the naked haircells. At low Reynolds numbers, the drag force F exerted on a prolate shaped cupula can be written as F = C.μ.L.U, where C is a constant, U is the flow velocity, L is the characteristic length or the diameter of the structure facing the flow, and μ is the dynamic viscosity. Using a simple scaling analysis, the ratio of the drag force exerted on a hydrogel cupula as compared to the naked haircell sensors can be estimated using a prolate shape approximation, as follows

$$\left(\frac{F_{cupula}}{F_{hair}}\right) \approx \left(\frac{H_{cupula}}{H_{hair}}\right)^{4/3} \left(\frac{D_{cupula}}{D_{hair}}\right)^{2/3} \qquad (1)$$

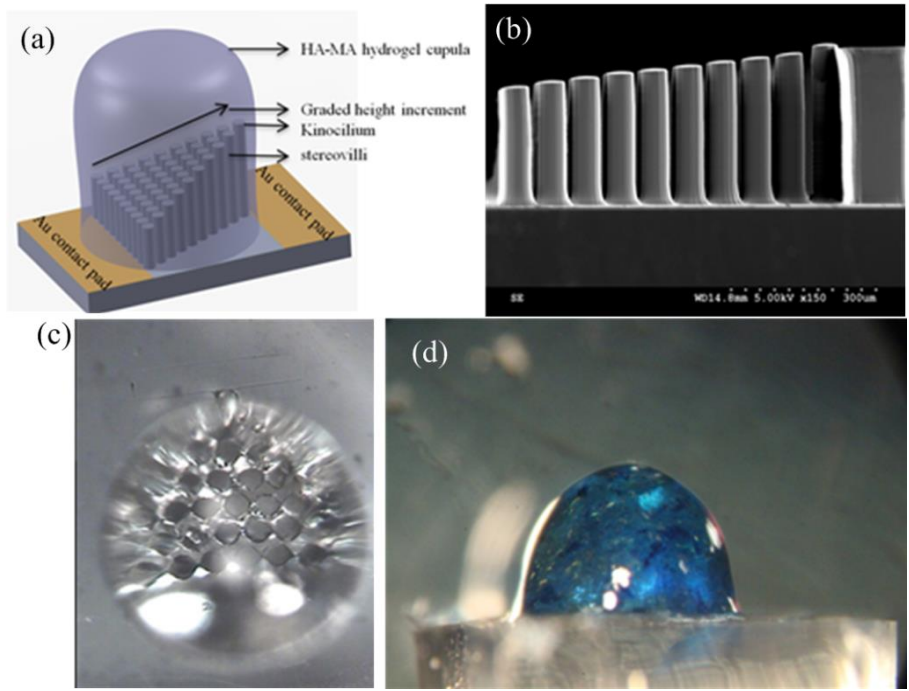

Figure 2. (a) A schematic describing the artificial stereovilli device structure and the various materials used in the device fabrication. (b) SEM images of the artificial haircell bundle consisting of 55 SU-8 stereovilli with varying aspect ratios (c) & (d) show optical microscopic images of the artificial haircell with flexible PDMS pillars and HA-MA hydrogel cupula formed on the haircell bundle.

Using the dimensions of the cupula and the haircell, the drag force due to the presence of the cupula is 2.2 times higher than for the naked stereovilli. The height of the haircell was taken to be the average height of all the 55 stereovilli and the diameter of the haircell was taken to be the distance between the shortest haircell row and the longest kinocilium.

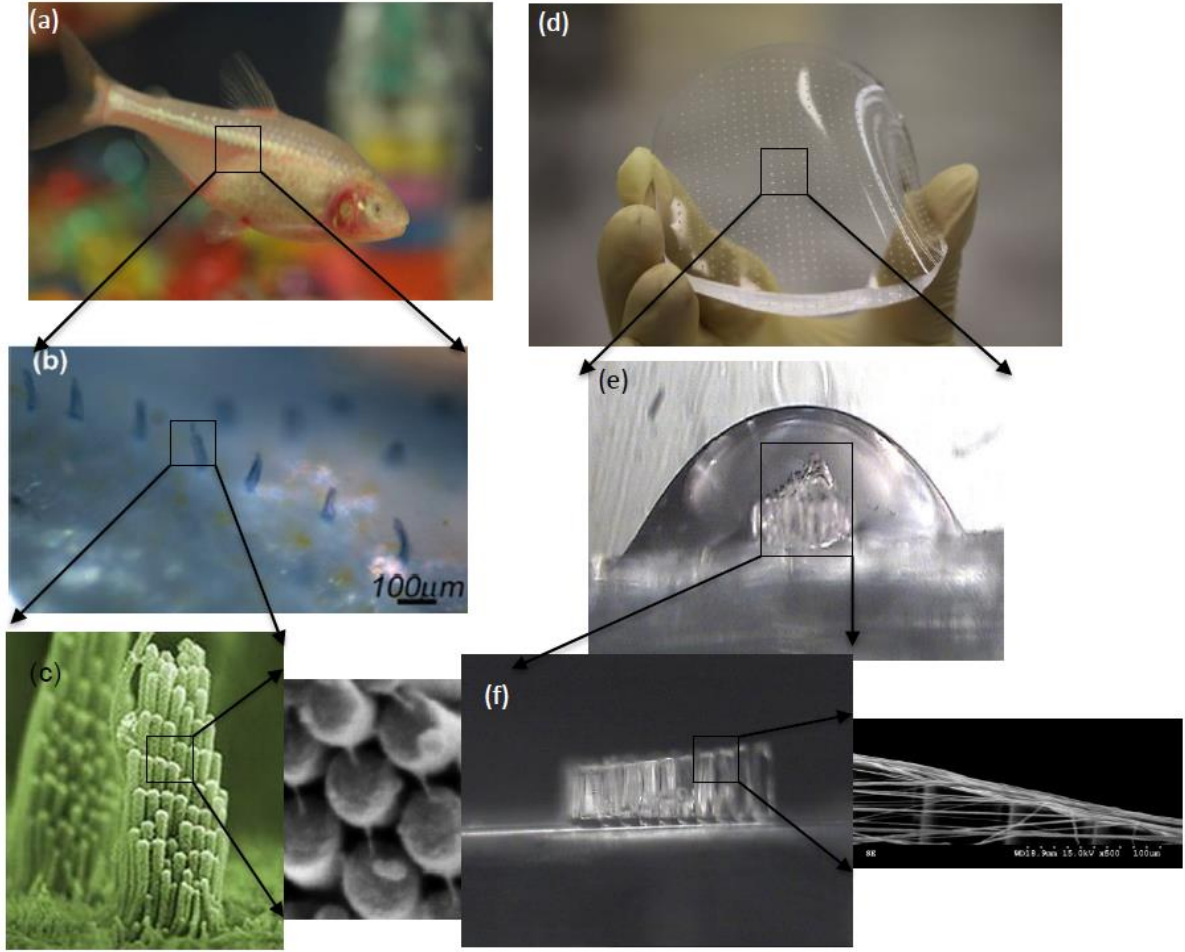

Figure 3. A series of photographs that draw an analogy between the structural features of the biological stereovilli of a single haircell bundle and the MEMS artificial stereovilli sensor. (a) A photograph of the blind cave fish which uses thousands of such haircells for flow sensing; (b) A photograph of a series of superficial neuromasts [1]; (c) An SEM image of an individual haircell bundle showing the stereoville [1]; the insert gives a closer look at the or tip links connecting the stereovilli; (d) A flexible PDMS wafer with thousands of MEMS flow sensors; (e) An optical microscopic image of a single MEMS biomimetic superficial neuromast flow sensor; (f) A closer view showing the PDMS stereovilli encapsulated within the hydrogel cupula. The insert shows the nanofiber tip links that connect the distal tips of the PDMS stereovilli.

The sensor operates through the bending of the artificial cupula in response to flow disturbances, resulting in the bending of the embedded PDMS stereovilli. Due to the graduation in height of the pillars in the rows of the bundle, there is a difference in the displacement of the pillars' tips in the various rows and, therefore, there will be a stretch induced on the PVDF nanofibers connecting the pillars' tips. This stretching occurs at all the nanofiber tip links. The charges generated in the nanofibers are collected at both ends through contact pads as a voltage output. Figure 4a provides a schematic of the bending in stereovilli of the same height, as compared to that of stereovilli with graduated height. This is a two-dimensional representation, each pillar representing an array of pillars.

Figure 4a explains why a graduated height is beneficial for generating a strong sensor output. When the direction of flow is along the direction of increasing height of the stereovilli, the deflection of the shorter pillars is towards their taller neighbours causing a relative shearing motion that results in stretching and tension increase at the tip links (flow direction is $0°$). This is the most sensitive axis of the MEMS flow sensor, which causes a maximum stretching of the nanofibers, thereby resulting in a maximum sensor output. In the case of $180°$ flow direction, the longer pillars bend toward their shorter neighbours, causing the nanofiber connecting the pillars to relax. This is the least sensitive case, wherein the fibers do not generate any output. Intermediate flow directions between 0° and 180° result in sensor outputs that can be approximated by a cosine function of the incident flow direction.

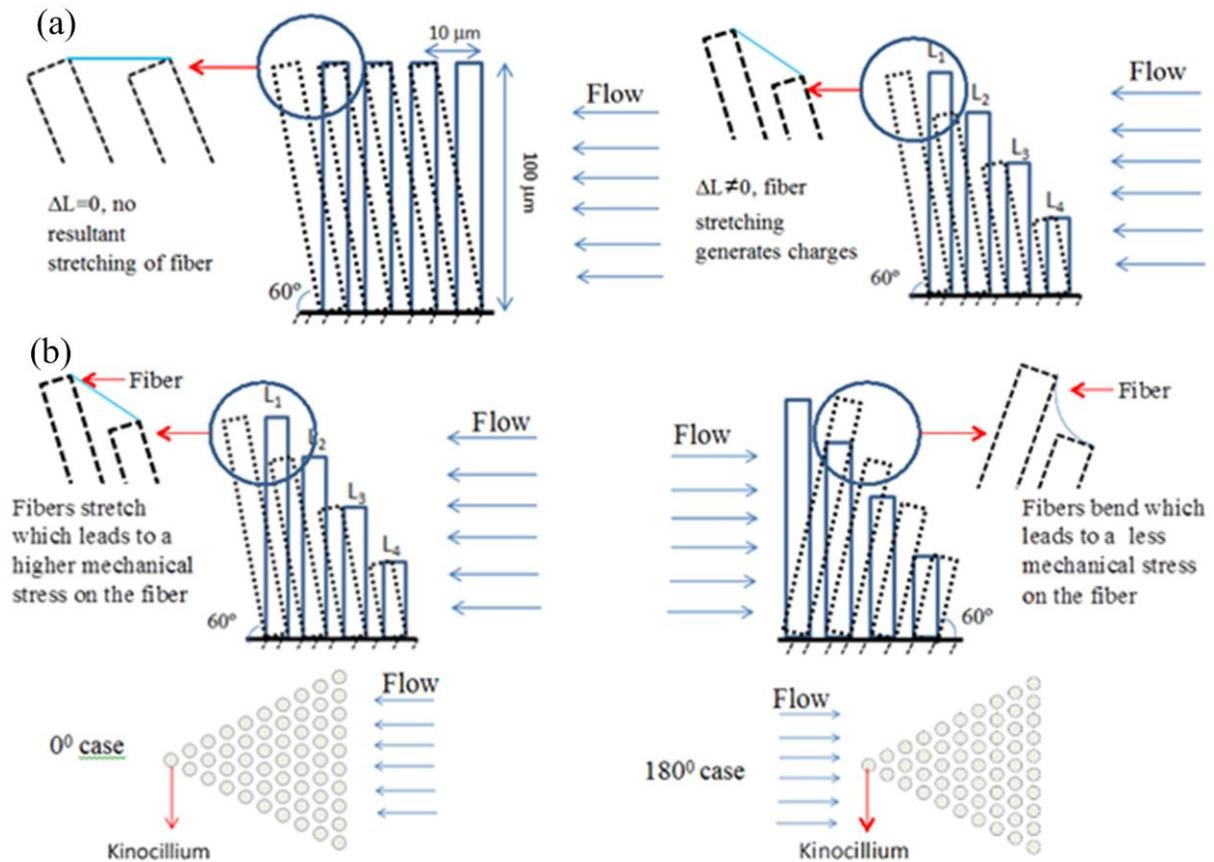

Figure 4. A schematic describing the basic sensing principle of the biomimetic MEMS sensor. Each pillar in the 2D schematic represents all the pillars in a particular row of the same height (the rows extend perpendicular to the plane of the paper). (a) If pillars of all the rows are designed to have the same height, there would be no resultant stress generated on the nanofibers (b) Pillars with height gradient result in stress in each unit tip-link that connects a pillar to its successive taller pillar. (b) Directional sensitivity of the biomimetic MEMS flow sensor (a) $0°$ flow direction causes the pillars to bend towards the longest kinocilium causing a maximum stretch in the fibers and a maximum sensor output. The sensor shows highest sensitivity along flows in this axis (b) $180°$ flow direction causes the pillars to bend away from the longest kinocilium causing a slacking of the fibers and resulting in zero sensor output. The sensor shows least sensitivity along flows in this axis

To evaluate the performance of the sensor, we performed three different experiments using a vibrating sphere (dipole) stimulus, driven with a sinusoidal signal to generate oscillatory flows of various amplitude and frequency (figure 5). The characteristics of the flow generated by the dipole source, and the reason for employing a dipole source, are illustrated in the supporting document.

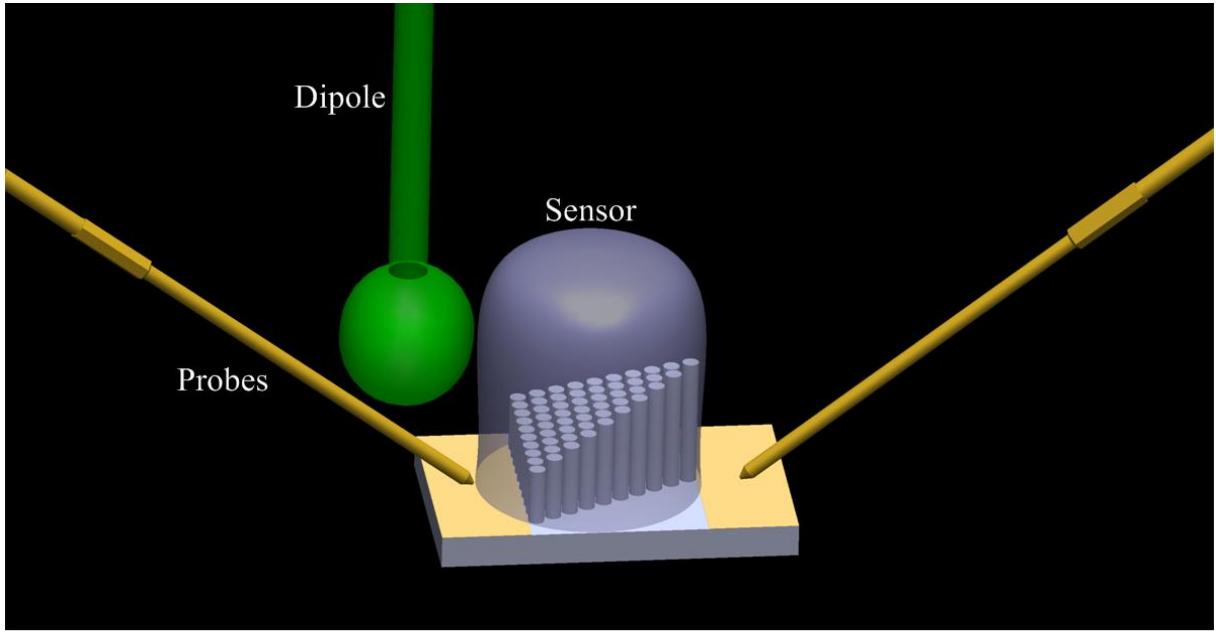

Figure 5. Experimental set-up in water showing probes that connect to the contact pads of the stereovilli sensor and the dipole in the vicinity of the sensor.

In the first experiment, the dipole was driven with a sinusoidal voltage at frequencies of 2 Hz and 35 Hz and constant amplitude of 250 mV$_{rms}$, while the sensor output was amplified by a gain of 500. The dipole was positioned at a distance of 25 mm away from the sensor and was vibrated in a direction parallel to the long-axis of the stereovilli. The vibration of the dipole causes the water pressure in the vicinity of the dipole to fluctuate at the same frequency, forcing the cupula to bend, and resulting in the bending of the stereovilli as described above, hence generating electric charges. Figure 6a and 6b show the output of the sensor to dipole vibrating at 2 Hz and 35 Hz.

The goal of the second experiment was to determine the sensitivity and velocity detection threshold of the sensor. The flow velocity was changed from 1 μm/s to 80 mm/s by adjusting the vibration of the dipole. The dipole was vibrated at constant frequency of 35Hz and the amplitude of vibration was varied in order to generate varying flow velocities. The sensor was positioned at a distance of 25 mm from the dipole and the output of the sensor was amplified by 500 times using SRS low-noise pre-amplifier. The flow velocity generated by dipole at the location of sensor is a function of sphere diameter, angular vibration frequency, frequency of vibration, displacement amplitude of dipole, and the distance between the center of dipole and sensors (observation distance) [24,32,33]. We conducted laser doppler vibrometer characterization of the dipole to determine the velocity of the dipole and applied the model developed in [34] to estimate the flow velocity generated by dipole at the location of the sensor.

Figure 6*c* shows the flow velocity calibration of the sensor. The results presented are average results by performing the same experiment on 5 sensors. The sensor's output varied linearly with respect to the amplitude of the sinusoidal source, signal as expected. The sensors demonstrated a threshold sensing limit of 8.24 μm/s, below which the sensor's response approached the noise limit. A high sensitivity of 0.286 mV/(mm/s) was achieved during the flow velocity sensing experiments. In figure 6c, it can be observed that the sensor output increases with increasing flow velocity. However, there is a transition in the increasing trend around flow velocities 30-40 mm/s. This is not due to the performance of the sensor, but due to the mechanics of flow-structure interaction. The Reynolds number ($R_e$) for flows at this transition point is 50-70. Skin friction dominates in contribution to sensor output at low flow velocities below this $R_e$. However, at higher velocities, beyond 30-40mm/s), the contribution of pressure gradient dominates. While drag force due to skin friction is linearly proportional to flow velocity, drag force due to pressure gradient is quadratically proportional to velocity which causes a steeper increase in sensor output at velocities beyond 30-40mm/s.

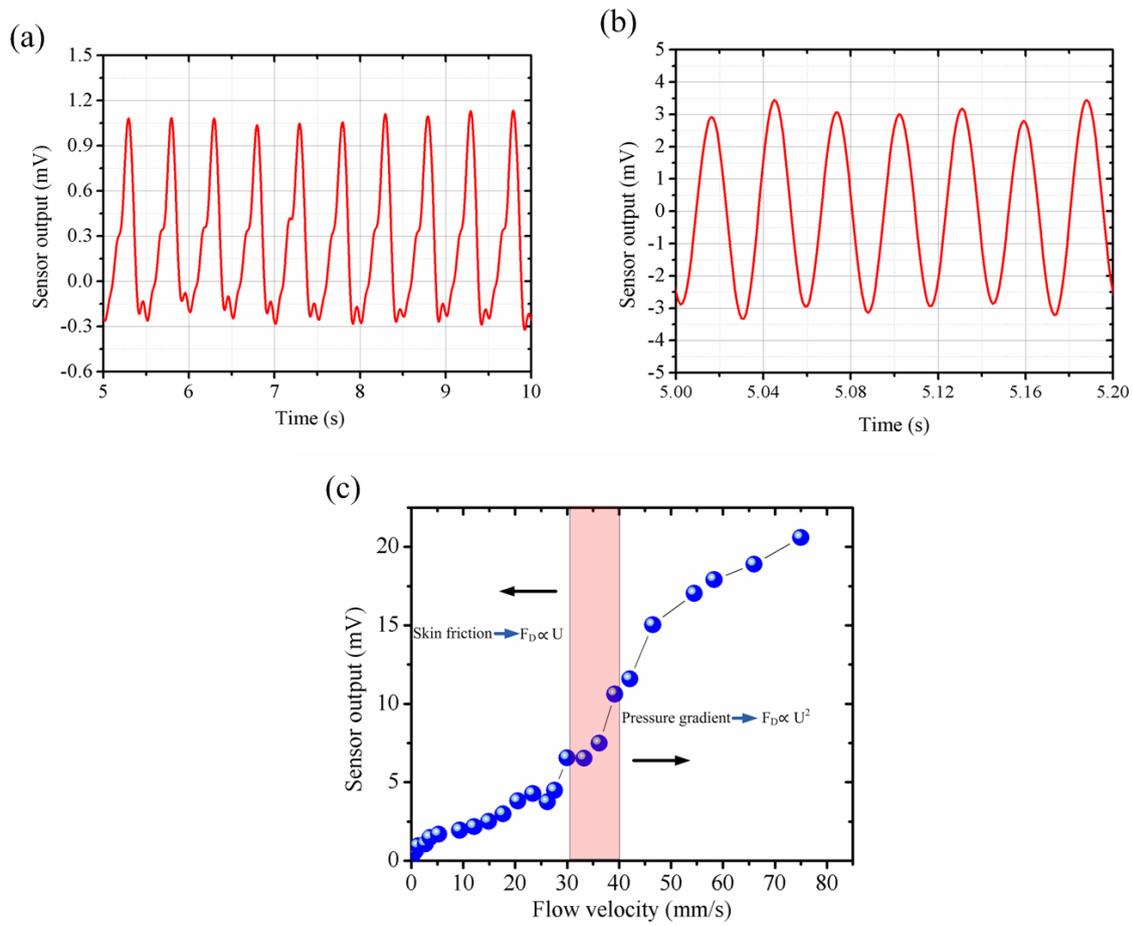

Figure 6. Sensor outputs for dipole stimulus vibrating at (a) 2 Hz and (b) 35 Hz. (c) Flow velocity calibration of the MEMS stereovilli sensors conducted in the water tunnel.

In order to determine the directional dependence of the sensor output, the position of the dipole was shifted at various angles with respect to the MEMS haircell bundle. Figure 7 shows the various directions tested.

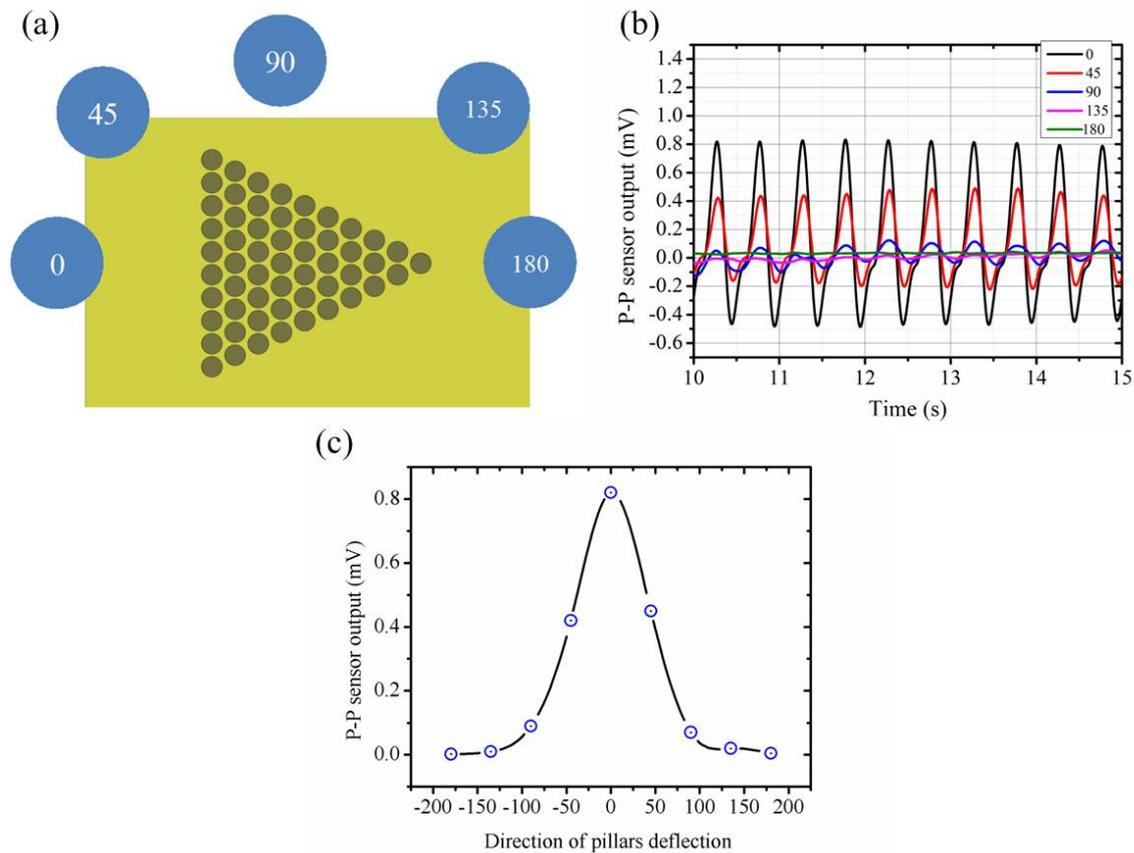

Figure 7. Directional dependence of the sensor output. (a) Schematic describing the various orientations of the dipole with respect to the stereovilli sensor (b) and (c) The sensor outputs for various orientations denoted in (a).

The sensor shows a large variation in its output as the dipole was shifted from $0^0$ to $180^0$. The best performance was for $0°$ when the sensor showed a maximum voltage output, because the nanofiber tip links were stretched to the maximum extent. At $180°$ the sensor generates almost no output since the mechanical stress on nanofibers the nanofibers is minimum. In directions between $0°$ and $180°$, the sensors output varies as a cosine function of the orientation. These results match very well with the predictions on the direction detection ability of the biological stereovilli sensors [14].

## Conclusions

An artificial MEMS ciliary bundle that mimics the structural features of the superficial neuromast in fishes is developed, built, and tested. The sensor features arrays of micro-pillars with gradated height (55 PDMS pillars organized into 10 rows) that are connected with electrospun nanofibers at their distal tips. A HA-MA hydrogel cupula is developed through a drop-casting process that encapsulates all the stereovilli and their nanofiber tip links. Since all the PDMS stereovilli are infused into the hydrogel cupula, when the cupula bends in response to an external flow, it causes all the stereovilli to bend as well. The difference in height of the pillars between rows causes a difference in the displacement of their tips, which leads to stretching and a resultant stress on the PVDF nanofiber that connects the pillars' tips. Water flow sensing experiments conducted using a dipole stimulus demonstrate a high sensitivity of 0.286 mV/(mm/s), and a low threshold detection limit of 8.24 μm/s.